# Architectures of Virtual Decision-Making: The Emergence of Gender Discrimination on a Crowdfunding Website
*Jason Radford, University of Chicago*

## 1. Introduction

The increasing relevance of Internet-based markets requires a sustained investigation into the relationship between design and user behavior. This research begins within the sociology of quantification and markets to investigate the impacts of basic design decisions on user behavior and individual success on a widely used crowdfunding website. This study looks at one common design feature, publishing recipients' sex, on the probability of receiving funding. Following research in the sociology of gender, these effects are defined along individual, behavioral, and structural dimensions. The results reveal that before teachers' sex was published, gender discrimination was weak and inconsistent. However, afterward gender discrimination increases by an order of magnitude and becomes systematized. Contrary to expectation, donors did not discriminate by sex category, but by teachers' structural position and the kinds of language they used. Implications for research on gender discrimination, priming, and online behavior are discussed.

## 1.1 Virtual Markets and the Architecture of Decision-Making

Virtual markets are becoming increasingly prevalent: matching workers to work, individuals to jobs, seed funding to entrepreneurs, liquidity to businesses, and philanthropists to causes. As sociologists of finance have pointed out, these internet-based markets are distinguishable from traditional, in-person marketplaces, in part, by their all-to-all, scopic architecture replacing traditional networks of direct exchange (Cetina and Bruegger 2002; White 1993). Research on this shift in the socio-technical structure of markets has demonstrated the relative importance of these new information architectures for local decision-making (Saavedra, Hagerty, and Uzzi 2011; Beunza and Stark 2011).

Simultaneously, social scientists have become interested in new forms and features of virtual social organization such as crowdsourcing, virtual social structures, and distributed cognition and human computation (Bainbridge 2007; Watts 2007). This research has shown how communities organize themselves by adapting their technological architecture to allocate and manage work. Thus far, little research has attempted to examine how the architecture of virtual markets influences their social organization. For example, while crowds have long been seen as calculators (Galton 1907); studies on the ramifications of virtualization for the process and outcomes of market resource allocation have been sparse (see Litzenberger, Castura, and Gorelick 2012 for exceptions in finance). This research attempts to bridge these by providing evidence for the relationship between information and decision-making in virtual a market. In the present case, a slight change in information rather than organization, generated a new, systematic allocation of resources among a national market of teachers in the United States.

## 1.2 Gender and Decision-Making

Existing research has demonstrated that providing individuals with gendered information changes the way they process information, make choices, and behave; most often in ways that typically advantage men or disadvantage women. However, while existing research suggests that adding information regarding teachers' sex should produce some discrimination, questions surrounding the reproducibility of priming effects, the limited range of gender information typically used, and the simplicity in how the effect is analyzed call into question whether simply adding "Mr." or "Mrs." to teachers' names would change donors' behavior (Cameron, Brown-Iannuzzi, and Payne 2012; Doyen et al. 2012). The data from DonorsChoose provide a strong, externally valid test for these effects that can be captured in heretofore under-explored dimensions.

Following the gender system perspective, this research defines gender along three dimensions (Risman 2004). First, the individual dimension of gender relates to beliefs about and categorization within the male/female binary (Ridgeway and Correll 2004). Second, the interactional dimension of gender refers to how one behaves in interaction as it relates to the binary, that is, whether one behaves in masculine or feminine ways (West and Zimmerman 2009). Thirdly, the institutional dimension refers to gendered expectations for behavior embedded in social structures whether institutions, organizations, or common social contexts (Acker 1990; Williams 2000). Due to sparse data, the need for experimental simplicity, and difficulty distinguishing the dimensions empirically; most research focuses on only one of these dimensions and few studies have investigated the interaction between them. This study examines all three simultaneously to investigate the extent to which donors discriminate and whether and to what extent this discrimination changes when new information is provided within the virtual market.

## 2.0 Data and Methods

DonorsChoose is a crowdfunding website for public school teachers in the United States. Since its inception in 2000, over 90,000 teachers from more than 40,000 schools have posted almost 300,000 projects. The data for all of these have been provided in this analysis. However, only data from between 2005 and 2010 are used as the event of interest, publishing teachers' sex, occurred in early 2008 as identified by archived versions of the website available on the Internet Archive. In the final analysis, a single linear model predicting the probability of projects' success is compared before and after teachers' sex is published. For this model, the three dimensions of gender are operationally distinguished from one another and specified as independent and interacting paramters.

The three dimensions of gender are operationalized using a combination of regression and text classification methods. Individual gender is a binary variable taken directly from teachers' self-reported sex; "Mr.", "Mrs.", and "Ms."; which teachers reported to DonorsChoose throughout its history. In all, over 99.9% of teachers provided a "Mr." "Mrs." or "Ms." since 2005 and those reporting nothing or "Dr." are excluded. Interactional gender, that is, how teachers behave, was inferred from the essays teachers wrote to appeal to donors. Naïve-Bayes classification; chosen for its human readability, facility for external validation, and parametric simplicity; was used to measure the relative "masculinity" or "femininity" of individual essays. The measure used was the classifier's likelihood ratio of a text being written by a "Mr." (masculine) rather than written by a "Mrs." or "Ms." (feminine). Lastly, institutional gender was operationalized with a linear model using structural features to predict the probability a teacher was male. Specifically, the model used grade-level and subject taught, whether a teachers' school was traditional or non-traditional school, and the school's metropolitan type to predict the binary measure of sex. In essence, institutional gender is the proportion of teachers expected to be male given a particular set of structural characteristics. In summary, individual gender measures whether or not a teacher is male or female. Interactional gender measures whether or not an essay is written in a masculine or feminine way. And, institutional gender measures the extent to which a teacher is more likely to be a man or woman given their school, grade, and subject. The crux of the analysis relies on using these three measures of gender to predict the probability that a project will receive funding before and after sex was published.

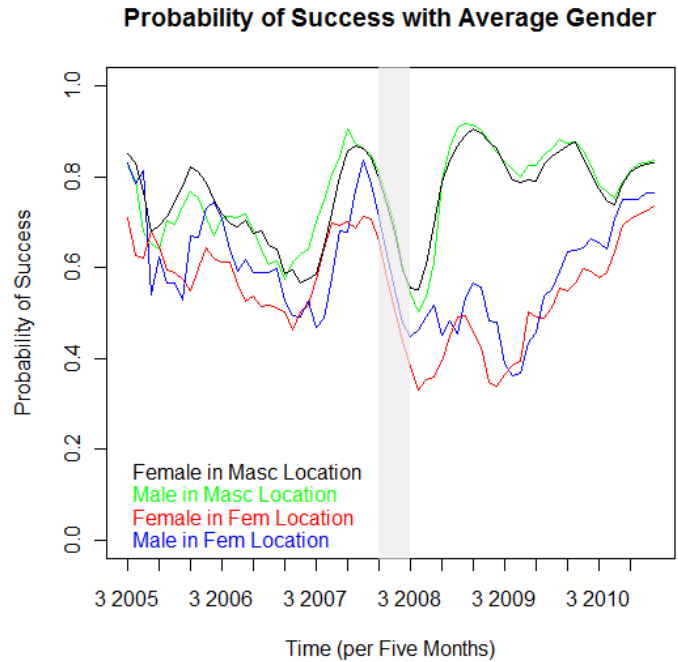

Figure 1: This chart shows the scale of change in the probability of funding after sex was published (grey area). The figure is produced by re-estimating the linear model containing controls and the variables of interest on five months of data at a time starting in 2005. The x-axis is the median month of the window while the y-axis is the predicted probability for a specific combination of individual and institutional sex. As the graph shows, after sex was published, both male and female teachers in male-dominated locations (black and green lines respectively) became much more likely to get funding than men and women in female-dominated locations (blue and red lines).

3.0 Results and Discussion

The results of the analysis show the emergence and systematization of bias, providing new insight into the social structure of gender. First, as social psychologists have argued, sex category appears to be a sparkplug, turning gender on. However, the differentiation it initiates is a differential evaluation of institutional expectations. It's the gendering of the social context (grade, subject, and school) that constitutes the largest source of discrimination while the way teachers actually behave has a weaker, but significant influence.

Several alternative hypotheses were tested. First, the financial crisis occurring around this time largely occurs after discrimination has already developed. Second, the influx of new teachers are compared separately and results show both pre-existing and new teachers had the same probability of receiving funding. Finally, there appear to be no network effects in predicting gender discrimination. However, data limitations prevent a full test of these alternatives. Indeed, *some* of the effects are extinguished by 2010, but tests of different hypotheses fail to explain this.

## 4. Conclusion

The results of this study not only push the sociology of gender forward by elaborating the relationship between the three dimensions of gender, but also demonstrate the power of information for social structure in virtual markets. It's most notable that information about teachers' sex was implemented as passive information which could not be used to navigate the website whether by searching, sorting, or filtering projects. It was also published in only a few places on an otherwise already well developed and widely used website. Yet, when this piece of information was added, a systematic gender evaluation process was initiated and millions of dollars reallocated along gendered lines.